\documentclass[useAMS,usenatbib]{mn2e}
\usepackage{epsfig}
\usepackage{float}
\usepackage{url,bm,color}
\def\Pm{\mbox{\rm P}_M}
\def\Rm{\mbox{\rm R}_M}

\def\Rey{\mbox{\rm Re}}
\newcommand{\EQ}{\begin{equation}}
\newcommand{\EN}{\end{equation}}
\newcommand{\EQA}{\begin{eqnarray}}
\newcommand{\ENA}{\end{eqnarray}}

\newcommand{\Fig}[1]{Figure~\ref{#1}}

\newcommand{\Figs}[2]{Figures~\ref{#1} and \ref{#2}}

\newcommand{\Tab}[1]{Table~\ref{#1}}

{}
{}
{}

{}
{}
{}
{}
{}
{}
{}
{}
{}
\newcommand{\meanBB}{\overline{\mbox{\boldmath $B$}}{}}{}
{}
{}
{}
{}
{}
{}
{}
{}
{}
{}
{}

{}
{}
{}

\newcommand{\meanB}{\overline{B}}

{}

{}
{}

\newcommand{\BB}{\bm{B}}

\newcommand{\uu}{\mbox{\boldmath $u$} {}}

\newcommand{\JJ}{\mbox{\boldmath $J$} {}}

\newcommand{\AAA}{\mbox{\boldmath $A$} {}}

\newcommand{\ff}{\mbox{\boldmath $f$} {}}

\newcommand{\FF}{\mbox{\boldmath $F$} {}}

\newcommand{\nab}{\mbox{\boldmath $\nabla$} {}}

\newcommand{\SSSS}{\mbox{\boldmath ${\sf S}$} {}}

\newcommand{\DD}{{\rm D} {}}

\newcommand{\dd}{{\rm d} {}}

\def\la{\mathrel{\mathchoice {\vcenter{\offinterlineskip\halign{\hfil
$\displaystyle##$\hfil\cr<\cr\sim\cr}}}
{\vcenter{\offinterlineskip\halign{\hfil$\textstyle##$\hfil\cr<\cr\sim\cr}}}
{\vcenter{\offinterlineskip\halign{\hfil$\scriptstyle##$\hfil\cr<\cr\sim\cr}}}
{\vcenter{\offinterlineskip\halign{\hfil$\scriptscriptstyle##$\hfil\cr<\cr\sim\cr}}}}}
\def\ga{\mathrel{\mathchoice {\vcenter{\offinterlineskip\halign{\hfil
$\displaystyle##$\hfil\cr>\cr\sim\cr}}}
{\vcenter{\offinterlineskip\halign{\hfil$\textstyle##$\hfil\cr>\cr\sim\cr}}}
{\vcenter{\offinterlineskip\halign{\hfil$\scriptstyle##$\hfil\cr>\cr\sim\cr}}}
{\vcenter{\offinterlineskip\halign{\hfil$\scriptscriptstyle##$\hfil\cr>\cr\sim\cr}}}}}
\def\Pm{\mbox{\rm Pr}_{\rm M}}
\def\Rm{\mbox{\rm Re}_{\rm M}}

\def\Rey{\mbox{\rm Re}}

\def\EK{E_{\rm K}}
\def\EM{E_{\rm M}}

\def\HM{H_{\rm M}}

\def\cs{c_{\rm s}}

\def\kf{k_{\rm f}}

\def\Brms{B_{\rm rms}}

\def\urms{u_{\rm rms}}

\def\half{{\textstyle{1\over2}}}

\newcommand{\be}{\begin{equation}}
\newcommand{\ee}{\end{equation}}

\title[A unified dynamo in helical turbulence]
{A unified large/small-scale dynamo in helical turbulence}
\author[Pallavi Bhat, Kandaswamy Subramanian, and Axel Brandenburg]
{Pallavi Bhat$^{1,2}$\thanks{E-mail:
palvi@iucaa.ernet.in}, 
Kandaswamy Subramanian$^1$\thanks{kandu@iucaa.in},
and Axel Brandenburg$^{3,4,5,6}$\thanks{E-mail:brandenb@nordita.org} \\
$^1$Inter University Centre for Astronomy and Astrophysics,
Post Bag 4, Pune University Campus, Ganeshkhind, Pune 411 007, India
\\
$^2$Department of Astrophysical Sciences and Princeton Plasma Physics Laboratory,
Princeton University, Princeton, NJ 08540, USA
\\
$^3$Nordita, KTH Royal Institute of Technology and Stockholm University,
Roslagstullsbacken 23, SE-10691 Stockholm, Sweden\\
$^4$Department of Astronomy, AlbaNova University Center,
Stockholm University, SE-10691 Stockholm, Sweden
\\
$^5$JILA and Department of Astrophysical and Planetary Sciences, University of Colorado, Boulder, CO 80303, USA
\\
$^6$Laboratory for Atmospheric and Space Physics, University of Colorado, Boulder, CO 80303, USA
}
\begin{document}


\pagerange{\pageref{firstpage}--\pageref{lastpage}} \pubyear{2012}

\maketitle

\label{firstpage}

\begin{abstract}
We use high resolution direct numerical simulations (DNS)
to show that helical turbulence can generate
significant large-scale fields 
even in the presence of strong small-scale dynamo action.
During the kinematic stage, the unified large/small-scale dynamo
grows fields with a shape-invariant 
eigenfunction, with most power peaked at small scales or large $k$, as in \citet{SB14}.
Nevertheless, the large-scale
field can be clearly detected as an excess 
power at small $k$ in the negatively polarized 
component of the energy spectrum for a forcing with positively polarized waves. 
Its strength $\meanB$, relative to the total rms field $\Brms$,
decreases with increasing magnetic Reynolds number, $\Rm$.  
However, as the Lorentz force becomes important,
the field generated by the unified dynamo
orders itself by saturating on successively larger scales.
The magnetic integral scale for the positively polarized waves, characterizing the small-scale
field, increases significantly from the kinematic stage to saturation.
This implies that the small-scale field becomes as coherent as possible 
for a given forcing scale, which averts the $\Rm$-dependent
quenching of $\meanB/\Brms$.
These results are obtained for $1024^3$ DNS with magnetic Prandtl numbers
of $\Pm=0.1$ and $10$. 
For $\Pm=0.1$, $\meanB/\Brms$ grows from about $0.04$ to about $0.4$ 
at saturation, aided in the final stages by helicity dissipation.
For $\Pm=10$, $\meanB/\Brms$ grows from much less
than 0.01 to values of the order the $0.2$.
Our results confirm that there is a unified large/small-scale dynamo
in helical turbulence.
\end{abstract}

\begin{keywords}
MHD--dynamo--turbulence--galaxies:magnetic fields--Sun:dynamo--magnetic fields
\end{keywords}

\section{Introduction}

Astrophysical systems like stars and 
galaxies host magnetic fields that are coherent on the
scale of the system itself. These fields are thought to arise 
due to the action of a turbulent dynamo, whereby helical
turbulence combined with shear amplifies and maintains fields 
coherent on scales larger than the scales of random stirring.
Indeed, the scales of the stirring like convective scale in the Sun
or the supernova-induced turbulent scales in galaxies, are much smaller
than the coherence scale of the large-scale field.
A dynamo which amplifies fields on scales larger than that of the stirring 
is referred to as a large-scale or mean-field dynamo.

There are two major potential difficulties associated with mean-field dynamos.
One is that small-scale helical fields which are produced during mean-field
dynamo action, due to magnetic helicity conservation, 
go to quench the dynamo.
Thus, they have to be eliminated from the dynamo active region by some form
of magnetic helicity flux, to avoid such quenching \citep{BS05,BSS12,EB15}.
Equally important is the fact that, while mean-fields are being generated, the
same turbulence, for a large enough magnetic Reynolds number, $\Rm$, 
also generically lead to the small-scale or fluctuation dynamo.
The fluctuation dynamo rapidly generates magnetic fields coherent on scales 
of the order of or smaller
than the outer scales of the turbulence, and
in principle, at a rate faster than the mean fields 
\citep{kaz,KA92,S99,HBD04,Schek04,BS05,TCB11,BSS12}.
The question then arises as to whether and how the mean-field
dynamo operates in the
presence of such rapidly growing magnetic fluctuations.

This issue was partially addressed by \citet{SB14} (hereafter SB14), 
through direct 
numerical simulations (DNS) and by analyzing the \citet{kaz} model, 
focusing on the kinematic
regime. They showed that in this regime, the magnetic energy spectrum 
grows as an eigenfunction, i.e. at each wavenumber $k$ the spectrum
grows with the same growth rate. Nevertheless, there is indeed evidence for this large-scale 
field generation in the horizontally averaged large-scale field, 
which can also be seen as excess power at small $k$ 
in one of the oppositely helically polarized components.
However, they also found that the relative strength of the 
large-scale or mean-field component compared to the rms field, in the kinematic
stage, decreases with increasing $\Rm$ like $\Rm^{-3/4}$ for 
larger values of $\Rm$. 
From both an analysis of the
Kazantsev model including helicity \citep{kaz,VK86,S99,BCR05}  
and the DNS, SB14 showed that this is a result of the
magnetic energy spectrum peaking on small resistive scales even in the presence
of helicity. 
If such a feature persisted on saturation, it would be difficult to
explain the prevalence of large-scale fields.
Of course, as the dynamo-generated field grows,
the Lorentz force will become important,
first at small scales, and saturate the growth of small-scale fields. 
It is then important to determine whether 
the mean-field dynamo can continue to grow 
large-scale fields as the fluctuation dynamo saturates.
And can this large-scale field become more dominant at saturation, 
independent of $\Rm$?
Our aim in this paper is to answer these questions.

For this purpose we have run DNS of magnetic field growth in 
helically driven turbulence in a periodic domain, with resolutions
up to $1024^3$ mesh points. These simulations are designed to adequately
capture the dynamics of scales both smaller and larger than the forcing scale,
and run from the kinematic regime to nonlinear saturation.  
The next section presents the simulations that we have carried out
to use for our analysis. Section 3 sets out the results of our analysis
to determine the evolution of both the large- and small-scale fields
generated by helical turbulence.
Section 4 presents a
discussion of these results and the last section, our conclusions. 

\section{Simulating large-scale dynamos}

To study the growth of the large-scale or mean-field dynamo in the presence
of a small-scale or fluctuation dynamo,
we have run a suite of simulations of helically driven turbulence using the 
\textsc{Pencil Code}\footnote{\url{https://github.com/pencil-code}} \citep{BD02,B03}.
The continuity, Navier-Stokes and induction equations are solved
in a Cartesian domain of a size $(2\pi)^3$ on a cubic grid with 
$N^3$ mesh points, adopting triply periodic boundary conditions.
The fluid is assumed to be isothermal, viscous, electrically conducting
and mildly compressible.
The governing equations are given by,
\begin{eqnarray}
&& \frac{\DD}{\DD t} \ln{\rho} = -\nab\cdot\uu, \label{eq: continuity} \\
&& \frac{\DD}{\DD t} \uu = -\cs^{2}\nab \ln{\rho} +
\frac{1}{\rho} \JJ\times\BB + \FF_{\rm visc} + \ff,
\label{eq: momentum} \\
&& \frac{\partial}{\partial t} \AAA = \uu\times\BB -\eta\mu_{0}\JJ.
\label{eq: induction}
\end{eqnarray}
Here $\rho$ is the density related to the pressure by $ P=\rho c_s^2$, where $c_s$ is speed of sound. 
The operator $\DD/\DD t=\partial/\partial t + \uu\cdot \nab$ is the
Lagrangian derivative, where $\uu$ is fluid velocity field.
The induction equation is being expressed in terms of the vector potential
$\AAA$, so that $\BB=\nabla \times \AAA$ is the magnetic field,
$\JJ= \nabla \times \BB/\mu_0$ is the current density
and $\mu_0$ is the vacuum permeability ($\mu_0=1$ in the DNS).
The viscous force is given as
$\FF_{\rm visc} = \rho^{-1}\nab\cdot2\nu\rho\SSSS$,
where $\nu$ is the kinematic viscosity,
and $\SSSS$ is the traceless rate of strain tensor with components
${\sf S}_{ij}=\frac{1}{2}(u_{i,j}+u_{j,i})-\frac{1}{3}\delta_{ij}\nab\cdot\uu$.
Here commas denote partial derivatives.
The forcing term $f=f({\bf x}, t)$ is responsible 
for generating turbulent helical flow.
The forcing is maximally helical as described in SB14.
In Fourier space, this driving force is transverse
to the wavevector $\bm{k}$ and localized in wavenumber space about
a wavenumber $\kf$. 
It drives vortical motions in a wavelength
range around $2\pi/\kf$, which will also be the
energy carrying scales of the turbulent flow. 
The direction of the wavevector
and and its phase are changed at every time step in the simulation
making the force $\delta$-correlated in time;
see \cite{Br01,HBD04,SB14} for details.
\begin{table}
\begin{center}
\caption{
Summary of runs discussed in this paper.
Here the values of $\urms$ are from kinematic phase,
whereas $\Brms$ and $\overline{B}$ refer to values from the saturation phase
after $\sim 624$ eddy turnover times for Runs~A, B, and C
and after $\sim 463$ eddy turnover times for Runs~D, E, and F;
the eddy turn over time is given by $1/(\urms \kf)$.
}\vspace{10pt}{\begin{tabular}{lcrccccr}
\hline
\hline
Run & $\Pm\!$ & $\;\Rm\!$ & $\urms$ & $\Brms$ & $\overline{B}$ & $\overline{B}/\Brms$ &$N^3\;\;$ \\
\hline
A & 0.1 & 330  &  0.135  &  0.085  &  0.033 &  0.38  & $1024^3$ \\ 
B & 0.1 & 160  &  0.130  &  0.092  &  0.046 &  0.49  & $256^3$ \\ 
C & 0.1 & 65   &  0.130  &  0.092  &  0.055 &  0.59  & $256^3$ \\ 
D & 10  & 3375 &  0.135  &  0.078  &  0.017 &  0.22  & $1024^3$ \\ 
E & 10  & 1575 &  0.126  &  0.072  &  0.019 &  0.27  & $256^3$ \\ 
F & 10  & 665  &  0.133  &  0.082  &  0.026 &  0.32  & $256^3$ \\ 
\hline
\label{runs}\end{tabular}}
\end{center}
\end{table}

For all our simulations, we choose
to drive the motions at a wavenumber $\kf=4$.
This choice is motivated by the fact that we wish
to resolve both the small-scale magnetic field structures in any turbulent
cell and at the same time include scales larger than the flow (with $k < \kf$). 
The strength of the forcing is adjusted so that the rms Mach
number of the turbulence, $u_{\rm rms}$ in the code (where velocity
is measured in units of the isothermal sound speed), is typically 
about 0.1.
This small value implies also that the motions are nearly incompressible.
We define the magnetic and fluid Reynolds number 
respectively by $\Rm= u_{\rm rms}/\eta \kf$ and $\Rey = u_{\rm rms}/\nu \kf$,
where $\eta$ and $\nu$ are the resistivity and viscosity of the fluid.
The magnetic Prandtl number is defined as $\Pm =\Rm/\Rey = \nu/\eta$.

For the simulations reported here we take $\Pm=0.1$ and $\Pm=10$;
the former value was used in most of the DNS in SB14. This was motivated by the fact that
for such small values of $\Pm$, the non-helical
small-scale dynamo was expected to be much
harder to excite \citep{Isk07}, which would then provide a better chance of 
seeing evidence for the large-scale field in the kinematic stage.
SB14 however found an efficient small-scale dynamo even for $\Pm=0.1$.
This is related to the fact that in SB14 the forcing wavenumber
was chosen to be $\kf=4$, while in the earlier work of \cite{Isk07}
it was between 1 and 2.
Furthermore, the disappearance of the small-scale dynamo for $\kf=1.5$
is related to the bump in the spectrum near the dissipation wavenumber,
which is known as the ``bottleneck phenomenon'' \citep{Fal94}.
In the nonlinear regime, however, this bump is suppressed by the
magnetic field and therefore the strength of the small-scale
dynamo is nearly independent of $\Pm$ \citep{Br11}.
As we would like to
compare with the kinematic results of SB14 and extend it to
the nonlinear regime, we consider first the
value of $\Pm=0.1$.
For both values of $\Pm$, we have studied a range of $\Rm$.
The fiducial simulation of a helical dynamo considered in this paper
has a resolution of $1024^3$ mesh points and
$\Rm=330$ (hence $\Rey=3300$), which we from now on refer to as Run~A. 
The other run with a similar resolution of $1024^3$ mesh points
has $\Pm=10$ and is referred to as Run~D.
A summary of different runs used in this paper is given in \Tab{runs}.

A quantity that was also evaluated in the DNS of SB14 and which helps 
to understand how the growth rate of the fluctuation dynamo changes as compared
to the mean-field dynamo is the growth rate as a function of $k$, $\lambda(k)$, 
of the magnetic energy residing in each $k$ defined in the following manner.
We first extract the evolution of magnetic energy at a given $k$, $M_k(t)$
from $\EM(k,t)$ and perform a running average with a suitable window, to smooth the evolution
curve, $M_k(t)$. This window is chosen such that the smoothing reduces the noise
sufficiently but at the same time does not produce any new false features 
in the curve.
Subsequently, to determine $\lambda$, we take the logarithmic derivative
given by
\begin{equation}
\lambda=\frac{1}{M_k(t)}\frac{\dd M_k(t)}{\dd t}.
\label{lam}
\end{equation}
We determine the $\lambda$ for each $k$, to obtain $\lambda(k)$.
\begin{figure}
\centering
\epsfig{file=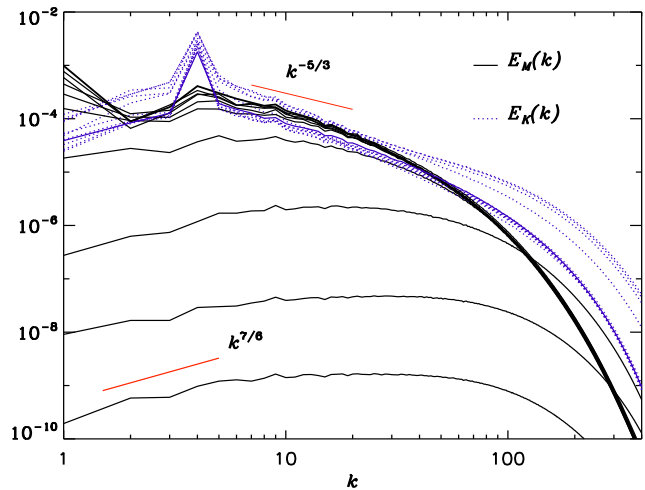, width=0.48\textwidth, height=0.27\textheight}
\caption{
Evolution of $\EM(k, t)$ for Run~A (solid black lines)
together with $\EK(k,t)$ (dotted blue), with the final curve
in solid blue. The spectra are at regular intervals of 100 code time units,
starting at $t=100$.}
\label{specLSD}
\end{figure}

\section{Results}

We first consider the results from one of our higher resolution simulations
of a turbulent helical dynamo with $1024^3$ mesh points (fiducial Run~A)
from the kinematic stage 
up to nonlinear saturation. Most of our results have
been obtained using the spectral data from the DNS.

\subsection{The evolution of spectra}

In \Fig{specLSD}, we show the evolving kinetic and 
magnetic energy spectra, $\EK(k)$ and $\EM(k)$, 
of the helically driven dynamo, 
in dotted blue and solid black lines
respectively. 
For a maximally helical velocity field, the 
mean-field $\alpha^2$ dynamo in a periodic box, 
is expected to grow fields initially at a wavenumber $k=\kf/2=2$ (see 
\citet{BDS02} and SB14),
which will then move to even larger scales (smaller $k$) in the saturated
state \citep{Br01}. On the other hand,
the fluctuation
dynamo is expected to be active at scales equal to and below the forcing scale 
(larger wavenumbers).
In the early kinematic stage, a single common eigenfunction is seen growing 
in a self similar manner, with 
the magnetic energy spectrum $\EM(k,t)$ increasing towards the smaller resistive
scales. This is similar to what is seen in the DNS of large-scale dynamo
action by SB14, during the kinematic stage.
Also the slope of $\EM(k)$ is close to $k^{7/6}$, which agrees with a
result derived by SB14 using Kazantsev model including helicity.

At late times, when the dynamo saturates,
the slope of $\EM(k,t)$ as a function of $k$
flattens first, with the peak of the spectrum shifting secularly to
smaller and smaller wavenumbers (larger and larger scales). 
The subsequent saturated spectra develop two peaks, one at the forcing scale
and another at the largest scale or smallest wavenumber, 
$k=k_1=1$. 

\begin{figure}
\centering
\epsfig{file=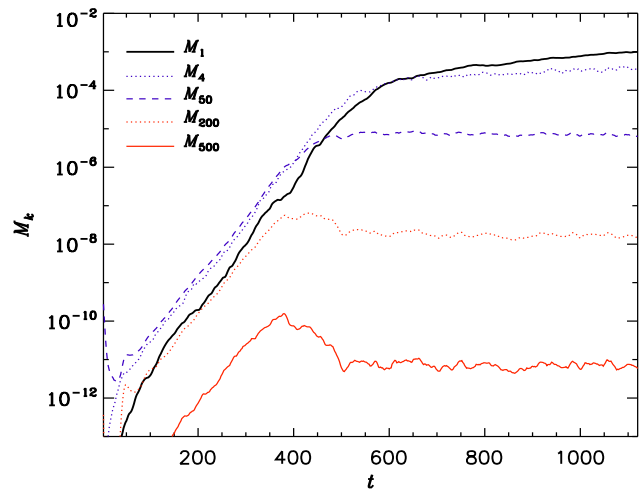, width=0.48\textwidth, height=0.27\textheight}
\caption{Evolution of $M_k(t)$ for $k=1$, $4$, $50$, $200$, and $500$ has been shown for Run~A.}
\label{diffktime}
\end{figure}

\begin{figure}
\centering
\epsfig{file=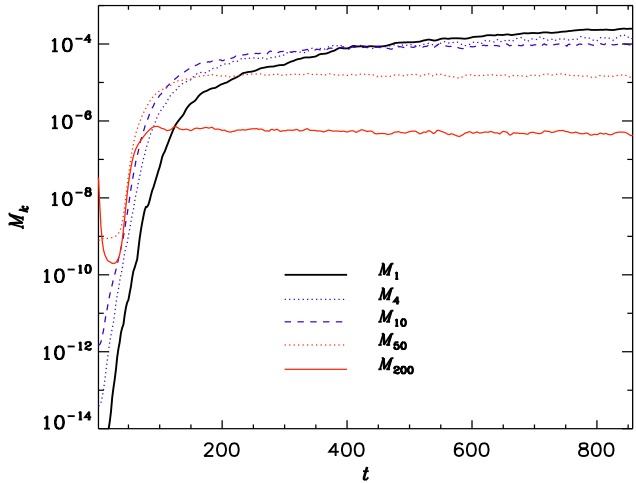, width=0.48\textwidth, height=0.27\textheight}
\caption{Evolution of $M_k(t)$ for $k=1$, $4$, $10$, $50$, and $200$ has been shown for Run~D.}
\label{diffktime2}
\end{figure}

\begin{figure}
\centering
\epsfig{file=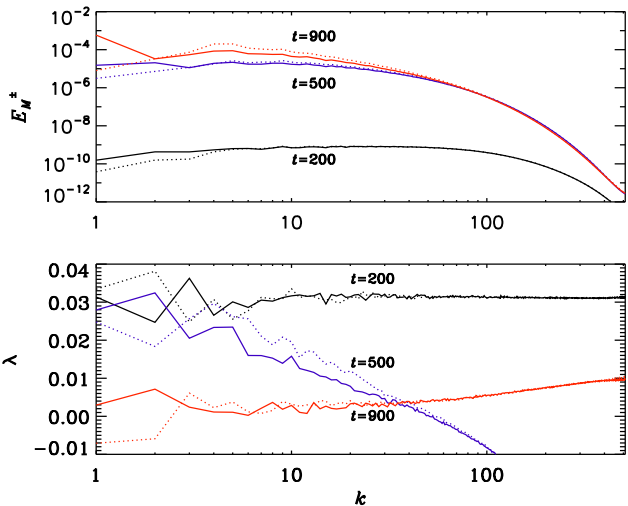, width=0.48\textwidth, height=0.28\textheight}
\caption{In the top panel, spectra of $\EM^{-}$ 
for Run A
are in solid curves and
$\EM^{+}$ are in dotted curves at three times, $t=200$, $500$, and $900$.
In the bottom panel the corresponding growth rate, $\lambda(k)$ is shown.}
\label{polsplit}
\end{figure}

To make this saturation behavior clearer, we show for Run~A
in \Fig{diffktime} the evolution of magnetic energy
residing in different scales.
The evolution of $M_k$ is shown for 
wavenumbers $k=1$, $4$, $50$, $200$, and $500$.
It can be seen that $M_k$ first grows almost exponentially 
with similar slopes for all $k$ in the kinematic stage,
before saturating. For 
higher wavenumbers, $M_k$ stops growing and turns to saturate
at lower strengths, and earlier than magnetic energy at lower wavenumbers. 
The $M_4$ and $M_{50}$ modes (dotted blue and dashed blue lines), 
turn to saturate at around $t=525$ and $450$, respectively, as compared to 
$M_1$ in solid black, 
which has not saturated even at late times. 
In fact, the $M_1$ mode, which reflects the operation of the large-scale dynamo,
is seen to be still growing, and has a distinctive positive
slope compared to the saturated flatter curves
at other wavenumbers.

A similar picture is seen also for Run~D ($\Pm=10$),
as shown in \Fig{diffktime2}. Here, the growth rate is much larger than for Run~A ($\Pm=0.1$).
This is because for $\Pm>1$ (and also large $\Rm$),
firstly, the smallest eddies whose scale-dependent $\Rm$ is greater than 
the critical value for small-scale dynamo action, have a shorter turnover time.
Secondly, the larger value of $\Rm$ for Run~D (by an order of magnitude,
compared to the $\Pm=0.1$ run), could make the dynamo in that run 
more efficient -- even for a similar eddy turnover rate.
Thus, the helical dynamo in Run~D turns to saturate at a much earlier time compared to Run~A.
As is well known starting from the work
of \citet{Br01} (see also \citet{BS05,BSS12,EB15} for reviews), 
this growth of large-scale field requires small-scale
helicity to be lost from the system.
In the present context of a completely homogeneous dynamo with
uniform energy density of the large-scale field, such a loss is
purely due to resistivity, whereas more realistically it would be
aided by helicity fluxes out of the dynamo active region. Nevertheless, 
the magnetic field evolution reflected in \Figs{specLSD}{diffktime} 
goes to show that even after the fluctuation dynamo
saturates, the mean-field dynamo continues to grow large-scale fields, 
provided also that small-scale magnetic helicity can be lost. 
\begin{figure}
\centering
\epsfig{file=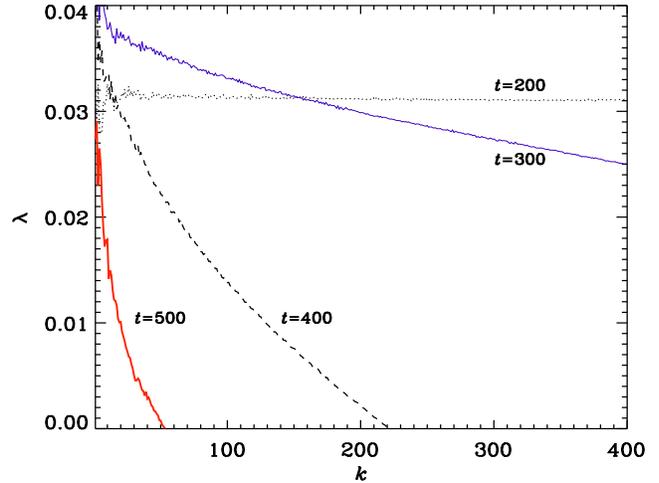, width=0.48\textwidth, height=0.27\textheight}
\caption{Growth rate $\lambda(k)$ for Run~A as a function of $k$ at
times $t=200$, $300$, $400$, and $500$.}
\label{lamk}
\end{figure}

An interesting behavior to note is that the $M_1$ mode
is growing together with all other modes at about the same rate and
turns off to saturate almost along with $M_4$. 
From this a picture emerges where
there is one unified large- and small-scale dynamo in such helical
turbulence \citep{S99}, which simply grows fields on all scales together,
and saturates at successively larger and larger scales (smaller
and smaller wavenumbers). We return to this idea below.

\subsection{Polarization spectra and wavenumber dependent growth rate}
\label{MagneticSpectra}

For the large-scale dynamo action which arises in helical turbulence,
the turbulent emf, in a two-scale model, 
is expected to generate oppositely signed 
small- and large-scale magnetic helicities.
Thus, one way of distinguishing large- and small-scale fields would be
to compute spectra from the field split into
positively and negatively polarized components, defined as
\citep{BDS02,BS05}
\begin{equation}
\EM^\pm(k,t)=\half\left[\EM(k,t)\pm\half k\HM(k,t)\right].
\label{Empol}
\end{equation}
This would enable one to see a clearer
signature of the large-scale field and its evolution from the
kinematic stage to nonlinear saturation.
In the top panel of \Fig{polsplit}, we show for Run A
the spectrum of the two oppositely polarized field 
components (depending on helicity), $\EM^\pm(k,t)$, 
at three times, $t=200$, $500$, and $900$.
The spectra of the negatively polarized field, $\EM^{-}(k,t)$,
are shown as solid lines and those of the positively
polarized field are shown as dotted lines. 
Also the continuous ordering of the field, and the continued growth of
large-scale field even when the small-scale field saturates, can also be
seen by examining the wavenumber-dependent 
growth rate, $\lambda(k)$. 
In the bottom panel of \Fig{polsplit},
the growth rates $\lambda(k)$ corresponding to $\EM^\pm(k,t)$ have been plotted.
The black curves in both panels are at $t=200$, in the kinematic stage.
We see that at this time $\lambda(k)$ is nearly uniform across $k$ in both
polarized components, indicating in the presence of an eigenfunction
which is growing at the same rate at all scales, both large and small.
(This can also be seen from \Fig{lamk}.)
We see that both polarized spectra extend over all $k$, 
and in fact overlap at large $k$, indicating that the magnetic field
has little helicity on such large $k$.  
However even at this time, there is an excess of power in the negatively
polarized field at small $k$, indicating the presence of 
a large-scale field due to the mean-field dynamo, as also 
found in SB14.

The blue curves at $t=500$, in the top panel, show that the 
negatively polarized (large-scale) field (given by the solid line)
near small $k$ is starting to rise above the rest of the spectrum.
Correspondingly we see from the bottom panel that the growth rate of the 
negatively polarized large-scale field component (solid curve) 
peaks at small $k$, whereas
that of the positively polarized field (the dotted line) peaks 
at around $k=4$. Furthermore, there is a decrease in
$\lambda$ for the rest of the spectrum.
This can also be noted from \Fig{lamk}, where the $\lambda(k)$ at $t=500$
for large $k$ had decreased significantly as compared to when $t=200$ or $t=300$.
Note that the peak in $\lambda(k)$ at $k=2$ for the negatively
polarized component, 
is consistent with the expectation that a fully helical
$\alpha^2$ mean-field dynamo grows fields at $k=\kf/2$.

Therefore, at these later times, 
there is no longer a growing eigenfunction.
Instead, the field is beginning to order itself both on large 
scales (due to the helical large-scale dynamo)
and also now on the forcing scale (due to the small-scale dynamo 
modified by the Lorentz force).
Finally, at $t=900$, shown as the red curves, the energy in the large-scale fields 
($\EM^{-}(k,t)$ given by the solid line) shows a peak 
at $k_1$ as seen in the top panel. In the bottom panel, 
we see that the growth rate
for both polarizations are close
to 0, but the negatively polarized large-scale field (solid curve)
still shows a positive $\lambda$ at small $k$. Thus, at this time the
small-scale field has saturated, but the large-scale field continues to grow
in the presence of a saturated small-scale dynamo 
(albeit due to the loss of helicity
through resistivity; see below).
\begin{figure}
\centering
\epsfig{file=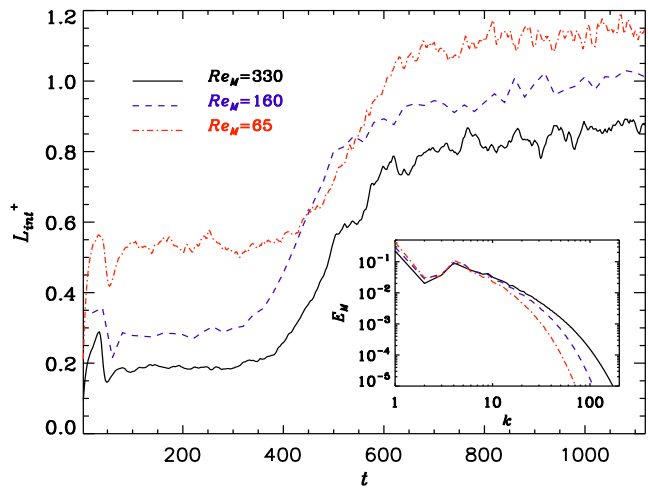, width=0.48\textwidth, height=0.27\textheight}
\caption{Evolution of the integral scale $L_{\rm int}^+$ for Runs~A, B, and C.
In the inset, their respective normalized saturated final magnetic spectra are shown.}
\label{ssint}
\end{figure}

\begin{figure}
\centering
\epsfig{file=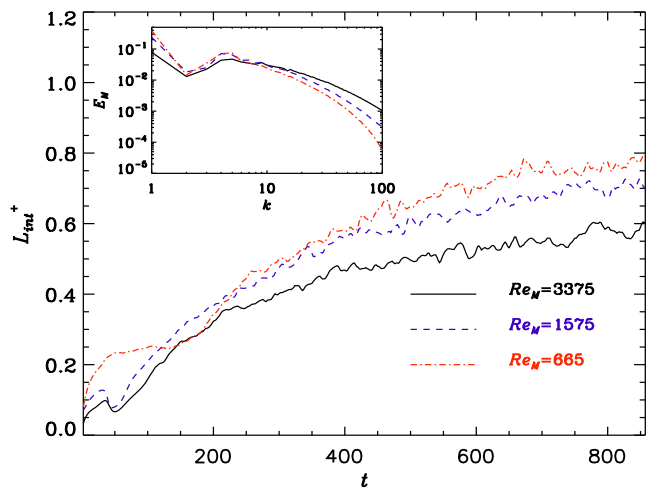, width=0.48\textwidth, height=0.27\textheight}
\caption{Evolution of the integral scale $L_{\rm int}^+$ for Runs~D, E, and F.
In the inset, their respective normalized saturated final magnetic spectra are shown.}
\label{ssint2}
\end{figure}

\subsection{Evolution of the small-scale field coherence}

As our work is focused on the evolution of
the large-scale field in the presence of growing small-scale fields,
it is also of interest to examine
the evolution of coherence properties of the small-scale field.
It is well known that the non-helically driven 
fluctuation dynamo generates fields
whose power is concentrated on resistive scales in the
kinematic stage. We have seen from our DNS (and from the work
of SB14) that this continues to hold even when the
turbulence is helical. 
In fact, in the kinematic stage, the spectrum is dominated by
power which is concentrated at resistive scales 
due to the fluctuation dynamo. For a high-$\Rm$ system, 
if such a feature persisted in saturation, the prevalence
of a mean field would be questionable. 
Thus, it is important to investigate whether for a high-$\Rm$ system, 
there is a shift of magnetic energy from resistive scales to
larger scales closer to the stirring scale on saturation.
To address this question, we show in \Fig{ssint}, the time evolution of
the integral scale of the positively polarized spectrum 
(which is predominantly the small-scale field), for Run~A.
Similar results were obtained if we define the energy spectrum of the
small-scale field to be $\EM(k,t)$ with $k>\kf$.
This integral scale, defined here separately for positively and negatively
polarized components, is given by
\begin{equation}
L_{\rm int}^\pm(t) =\frac{\int (2\pi/k) \EM^\pm(k,t)dk }{\int \EM^\pm(k,t)dk}.
\label{lint}
\end{equation}
In the following, we are particularly interested in $L_{\rm int}^+(t)$,
which characterizes the small-scale part of the field.
In \Fig{ssint}, the integral scale for the fiducial high resolution Run~A 
is shown as a solid black line.
We also show the results from two lower resolution runs,
Runs~B and C, with $\Pm=0.1$. 
In \Fig{ssint2}, we show the results from Runs~D, E, and F,
all of which have $\Pm=10$.

\begin{figure}
\centering
\epsfig{file=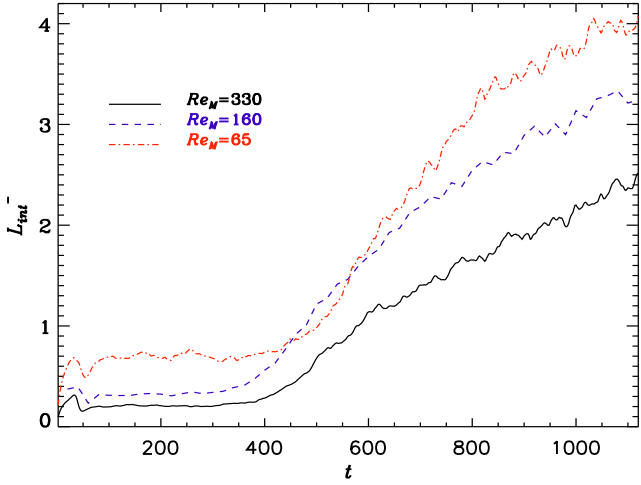, width=0.48\textwidth, height=0.27\textheight}
\caption{Evolution of the integral scale, $L_{\rm int}^-$, of the negatively
polarized $\EM^{-}(k)$ characterizing the large-scale dynamo, is shown for Runs~A, B, and C,
corresponding to $\Rm=330$, 160, and 65, respectively.}
\label{lintLS}
\end{figure}

For the fiducial run, we see that the integral scale is roughly 
constant at $L_{\rm int}^+\sim 0.17$
during the kinematic stage, reflecting the fact that the positively polarized 
field grows as an eigenfunction during this stage, with a small coherence scale.
However, as the Lorentz force becomes important, $L_{\rm int}^+(t)$ begins to
increase rapidly. This process begins at $t\sim 400$, which is also
the time when the large $k$ modes ($k> 50$) stop growing 
(see \Figs{specLSD}{diffktime}).
This rapid increase stops at $t \sim 600$ and $L_{\rm int}^+ \sim 0.8$, 
when the small-scale field modes with $k=\kf$ have largely saturated.
There is then a subsequent slower rise of $L_{\rm int}^+$ to $\sim 0.9$, 
as the large-scale field starts to dominate.
Thus, there is considerable increase (by more than a factor 5)
in the integral scale of the small-scale field from the kinematic
to the saturated state. 
This factor is higher than that of $\sim3$ seen for fluctuation dynamo in the
purely nonhelical case, albeit for $\Pm=1$ \citep{BS13}. 
In the case of $\Pm=10$, as for Run~D, $L_{\rm int}^+$ increases
from a value of $\sim 0.075$ in the kinematic stage to $\sim 0.6$ at saturation,
which is an increase by a factor of $\sim 8$.

Similar evolutions are also seen at lower $\Rm$. However here $L_{\rm int}^+$
is larger even in the kinematic state (reflecting the smaller resistive
wavenumber for a lower $\Rm$). And $L_{\rm int}^+$ also saturates
at a larger value for a lower $\Rm$, reflecting
the fact that the spectrum cuts off at smaller $k$ for lower $\Rm$.
This latter feature can be seen from the normalized spectra
shown in the inset of \Fig{ssint}.
Also note that, even though
the saturated $L_{\rm int}^+$ is slightly different for Runs~A, B, and C, in all three
cases, the peak power for the saturated small-scale dynamo is always at the forcing scale, $\kf=4$. 
In fact, one can also compare the $L_{\rm int}^+$ obtained for
our fiducial run with that expected for a small-scale
field coherent on the scale of forcing. Suppose we modelled
the spectrum of this field as $\EM^+ = M_0(k/\kf)^2$ for $k<\kf$
and $\EM^+ = M_0 (k/\kf)^{-5/3}$ for $k >\kf$, then one
gets $L_{\rm int}^+ =0.6 (2\pi/\kf)= 0.94$, which compares reasonably
well with that obtained in our highest resolution run.
Thus, it appears that, on saturation, the small-scale field at the
given forcing scale has become almost as coherent as possible.
Note that, if the power in the small-scale fields were still at resistive scales,
any peak in $\EM(k,t)$ at $k=1$ would have made negligible contribution
to $B_{\rm rms}$. Therefore, the above result provides some assurance that,
even in high $\Rm$ systems, the large-scale field can indeed be significant 
and reveal itself on saturation.

In the \Fig{lintLS}, we also show the evolution of $L_{\rm int}^-$ (for the large scale field)
for the runs with $\Pm=0.1$.
In the kinematic regime, these curves are only slightly higher in amplitude compared
to the corresponding $L_{\rm int}^+$ curves in \Fig{ssint}, thus indicating the slight excess in
energy at large scales.
Around the same time as $L_{\rm int}^+$, the $L_{\rm int}^-$ curves start increasing to higher values.
For Run~A, $L_{\rm int}^-$ increases from $~0.17$ to a value of $2.5$ which is about factor of $~3$
larger than the final $L_{\rm int}^+$.
The difference in the final values between the three curves for different $\Rm$ is because the
large-scale fields are still growing due to resistive dissipation of small-scale helicity.

\subsection{Growth of the large-scale field}

\begin{figure}
\centering
\epsfig{file=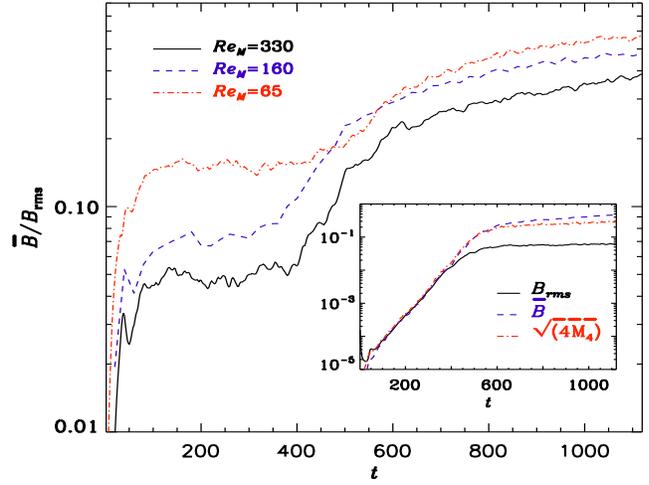, width=0.48\textwidth, height=0.27\textheight}
\caption{The ratio ${\bf \overline{B}}/B_{\rm rms}$ is evolving with time,
for Runs~A, B, and C. In the inset panel, for Run~A, the three
curves of $B_{\rm rms}(t)$, ${\bf \overline{B}}(t)$ and $\sqrt{kM_k}$ for $k=4$ are shown separately,
where the latter two curves are scaled by a constant to make all curves overlap in the
kinematic stage.}
\label{meanevol}
\end{figure}

\begin{figure}
\centering
\epsfig{file=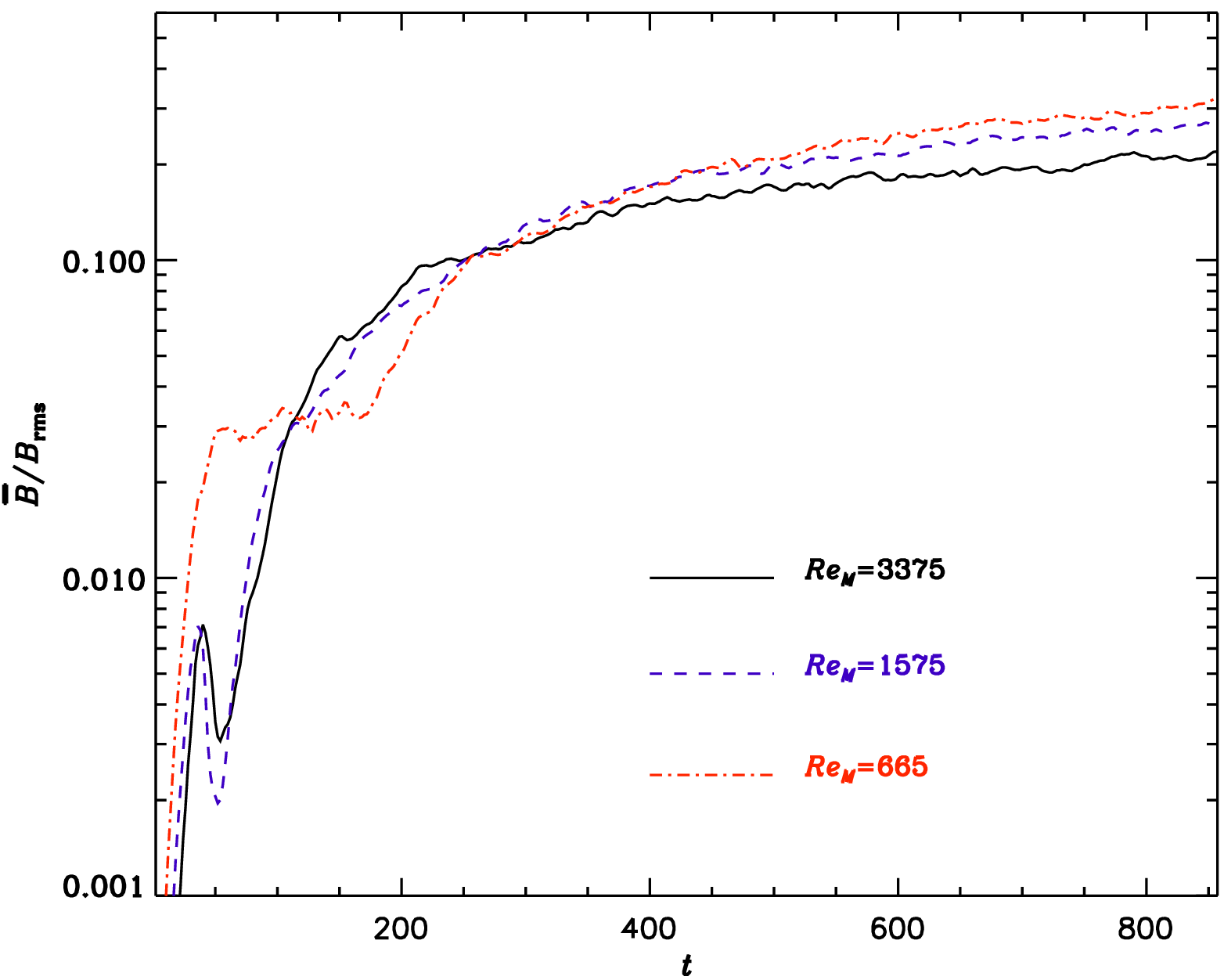, width=0.48\textwidth, height=0.27\textheight}
\caption{The ratio ${\bf \overline{B}}/B_{\rm rms}$ is evolving with time,
for Runs~D, E, and F.}
\label{meanevol2}
\end{figure}
We define the strength of the large-scale field $\meanB(t)$, 
by integrating the energy spectrum $\EM(k,t)$ between $k=1$--$2$ and
equating this to $\meanB^2/2$.
The ratio of strength of this large-scale field relative to the rms field
$\meanB(t)/\Brms$ is shown in \Fig{meanevol} evolving from the kinematic
stage to nonlinear saturation, for Runs~A, B, and C.
For the fiducial Run~A, shown as a solid black line, 
$\meanB/\Brms \sim 0.04$--$0.05$ in the kinematic stage. 
This is a factor 2 larger than $\meanB/\Brms$ determined by SB14 (Fig. 6)
on the basis of the mean power in horizontally averaged fields.
However we find that in the kinematic stage, the ratio $\meanB/\Brms$
in fact decreases with $\Rm$ approximately as $\Rm^{-3/4}$, 
which agrees with the scaling found by SB14 
for the horizontally averaged mean field.

The ratio $\meanB/\Brms$ begins to increase rapidly once the small-scale
field starts saturating, at $t\sim 400$ for Run~A.
This reflects the fact that, while the large $k$ modes saturate, the
$k=1$, $2$ modes which determine $\meanB$, continue to grow due to
large-scale dynamo action.
This can also be seen explicitly in the inset
of \Fig{meanevol}, where $\Brms$ in solid black turns to saturate at around $t=400$
while $\meanB$ in dashed blue continues to grow,
again reflecting the continued efficiency of the
mean (large-scale) field dynamo even when the fluctuation dynamo
begins to saturate. Finally, after $t=600$, this ratio enters a phase of 
slower growth; at this stage basically the large-scale field becomes
more and more dominant as the helicity of the small-scale field is lost
due to resistive dissipation. 
Note that, although larger $k$ modes begin to saturate by $t=400$ as was 
seen earlier also in \Fig{diffktime}, modes closer to the forcing scale (for example
the mode in $M_4$, also shown again in dashed red in the bottom panel of \Fig{meanevol}) 
are growing at the same rate as the modes in $k=1,2$ between $t=400$ to $500$
and they turn to saturate at nearly the same time, thus reinforcing the picture
of a unified dynamo. But from $t=600$ onwards, while the modes equal to $k=4$ 
and larger saturate completely, the $k=1,2$ modes continue to grow on resistive time
scales.
By the end of our run we have 
$\meanB/\Brms \sim 0.4$, for the fiducial run.
The lower $\Rm$ runs develop an even
higher value of $\meanB/\Brms \sim 0.5$ and $0.6$, for $\Rm=160$ and $65$, respectively.
This larger $\meanB/\Brms$ for lower $\Rm$ arises because the 
resistive dissipation of helicity is more important in these cases.

We show the evolution of the ratio $\meanB/\Brms$ 
also in the case of $\Pm=10$ in \Fig{meanevol2} for Runs~D, E, and F.
Here again the ratio $\meanB/\Brms$ increases dramatically from the kinematic
to the saturation regime.
We can assess how the scaling of the ratio $\meanB/\Brms$ vs.\ $\Rm^{\zeta}$
changes from the kinematic to the saturation regime.
In the former case of $\Pm=0.1$, $\zeta$ changes from a value of 
$\sim -0.75$ in the kinematic phase to $\sim -0.2$ during saturation.
In the case of $\Pm=10$, 
the kinematic phase is rather short in the current runs,
but one can see that 
$\zeta$ again tends to a small value of about $-0.2$ upon saturation.
The small residual $\Rm$ dependence of $\meanB/\Brms$ is also expected
to disappear in the final saturated state of the $\alpha^2$ dynamo,
where one expects this ratio to tend towards $(\kf/k_1)^{1/2}$, provided
the small scale helicity can be lost effectively \citep{Br01}. 
In the current simulations
as the helicity is only lost resistively, this has not yet occurred.
All in all, we see that a significant large
scale field can be generated even in the presence of an
active fluctuation dynamo. By the time they saturate
the curves come closer together indicating that the mean field
is dominant now.

\section{Discussion}
\label{Discussion}

Using high resolution DNS, we have shown
that, for helical turbulence at large $\Rm$,
the mean-field dynamo works efficiently to
generate significant large-scale fields --
even in the presence of a strong fluctuation dynamo.
It appears that there is only one unified large/small-scale dynamo
in such helical turbulence where initially fields on all scales 
grow together, and when the Lorentz force becomes important, 
successively larger scales saturate.

\subsection{Shape-invariant growth of the spectrum}
As in SB14, we find that in the kinematic stage 
the spectrum grows as a shape-invariant eigenfunction of the helical dynamo, 
peaked at small scales (or large $k$). 
There is clear evidence for a large-scale field even at this stage, 
seen as excess power at small $k$ in the negatively polarized 
component of the energy spectrum. 
However, the ratio of the strength of this large-scale field to the rms field
decreases with increasing $\Rm$.  This is due to efficient
fluctuation dynamo action, which amplifies power at small (nearly
resistive) scales.
The question then arises whether, in the presence of a fluctuation dynamo,
the large-scale field can grow to a significant fraction of the rms field
-- at least when the dynamo saturates.

\subsection{Scale-dependent saturation of the unified dynamo}

We show from the evolution of the spectra
(\Figs{specLSD}{diffktime}) that, as the field grows, 
small scales (large $k$ modes, $k>4$) saturate
first, but the large-scale field (with $k=1$) 
continues to grow at about the same rate as the $k=4$ mode, 
even when this happens. This can be seen
by examining the wavelength-dependent growth rate of both the differently 
polarized components (\Fig{polsplit}) and the total field (\Fig{lamk}).
These $\lambda(k)$ start out as being independent of $k$ in the
kinematic stage, but progressively decrease to zero, first at
large $k$ and then at smaller and smaller $k$. 
This saturation behavior where small scales saturate first
and then larger and larger scales saturate, remains qualitatively
unchanged for $\Pm=10$, even though the small scale
dynamo is more efficient, as can be seen in \Fig{diffktime2}.

\subsection{Increase of small-scale field coherence}

At the end of our simulation, the spectra displays two peaks, 
one at the forcing wavenumber $\kf$, and the other at $k=1$. 
Therefore the back reaction due to the Lorentz force has 
enabled the small-scale field coherence to increase from
small scales to the forcing scale, and at the same time
allowed the large-scale field to develop. 

The first feature can also be seen from \Figs{ssint}{ssint2},
where we show the evolution of the integral scale $L_{\rm int}^+$ of the positively
polarized component (identified with the small-scale field).
For our fiducial Run~A ($\Pm=0.1$), we show
that $L_{\rm int}^+$ evolves 
from a value of $\sim 0.17$ in the kinematic stage to
$L_{\rm int}^+ \sim 0.9$ upon saturation, a significant fraction of 
the forcing scale ($2\pi/\kf$).
Also in the case of $\Pm=10$, as in Run~D, $L_{\rm int}^+$
increases by a factor of $\sim 8$.
In fact, through nonlinear saturation, the small-scale field
has become as coherent as possible for the given forcing scale.

\subsection{Significant growth of large scale field upon saturation}

The growth of the large-scale field to significant levels, 
even in the presence of the fluctuation dynamo, was also shown
by considering the time evolution of $\meanB/\Brms$ 
(see \Figs{meanevol}{meanevol2}).
This ratio, in the case of $\Pm=0.1$, starts from a small value
of $\sim 0.04$ during the kinematic stage,
but at the end of our run, we obtain a significant large-scale field 
with $\meanB/\Brms \sim 0.4$. 
A large increase in this ratio in seen for also Runs~D, E, and F with $\Pm=10$.
The growth of the ratio in Run~A occurs in two stages:
First between $t\sim 400$--$600$ there is a rapid growth of $\meanB/\Brms$ 
as the fluctuation dynamo saturates. It appears that the nonlinear
ordering effects of the Lorentz force that saturate the small-scale fields,
still allow growth of progressively larger scale fields, including
scales larger than the forcing scale at $k < 4$. 
For $t > 600$, there is a slower growth of $\meanB/\Brms$ 
presumably driven by the resistive dissipation of 
the small-scale helicity. It would be interesting to ask
if, in very large $\Rm$ astrophysical systems, the effect of this resistive 
dissipation of small-scale helicity can
also be achieved by having instead magnetic helicity fluxes
\citep{BF00,Klee00}.

\subsection{Quantum mechanical analogy}

It may be instructive to think in terms of the Kazantsev model
incorporating helicity \citep{S99,BCR05,SB14}, 
where the dynamo problem is mapped to a quantum mechanical potential
problem, with growing small-scale dynamo modes mapped to bound states 
in the potential and with helicity allowing for tunnelling to have 
enslaved large-scale field correlations with the same growth rate.
The effect of the Lorentz force could be to make the
potential well at the small scale $\kf^{-1}$
to become shallower, allowing for only the marginally bound state to exist,
while still having sufficient depth at the large scale $k_1^{-1}$, 
to allow the ‘tunnelling free-particle’ states to grow.

Such behavior is indeed obtained
in a related real-space double-well potential problem,
arising in non-axisymmetric galactic dynamos, where 
the dynamo is enhanced along a spiral 
\citep{CSS13a,CSS13b}.
There the potential wells are near the galactic center and
the corotation radius of the spiral. The fastest growing
kinematic eigenfunction is largest in the central regions.
But its tail is enhanced along the magnetic spiral, near corotation radius.
However, saturation of the field near the galactic center,
still allows for the field to grow around corotation 
and become significant.
From our work here, it appears that such a situation can also be obtained
for a double-well potential in `scale' or wavenumber space, when
one incorporates nonlinear saturation effects. It would be of interest
to demonstrate this also in a nonlinear version of 
the Kazantsev model, perhaps generalizing the work of 
\citet{S99,BS00} to include helicity loss.

\subsection{The very limited role of $\alpha$ effect growth rates}

In the early days of mean-field dynamo theory, computing linear growth rates
was about the only way different dynamo modes could be characterized.
In the late 1980s, their limited usefulness became clear.
Only the marginally excited case of zero growth rates remained truly useful.
In particular, the dominance of one mode over the other is entirely
determined by nonlinearity, and not at all by differences in their kinematic growth
rates \citep{BKMMT89}.

Linear growth rates have traditionally also been used to estimate the
time it takes for the large-scale dynamo to reach saturation.
The linear growth rate of the $\alpha^2$ large-scale dynamo is expected to
be much smaller than that of the small-scale dynamo. We have seen however
that all modes grow together in the kinematic stage, and large-scale modes at
$k=1$--$2$ continue to grow at the same rapid rate even 
when the small-scale modes ($k>4$) saturate.
Note that it is the linear growth rate which has been important in 
discussions of the strength of the large-scale
field in young galaxies \citep{Kronberg,Bernet,JC13,FOCG14}.
Our present work does not really apply to galaxies, where shear is also
important and the magnetic Prandtl number is large, but it highlights
quite clearly that any estimate based on the value of $\alpha$, or
the value of $|\alpha\nabla\Omega|^{1/2}$ in models with differential
rotation $\Omega$ \citep{Beck96} must be irrelevant.
This was in principle already recognized by \cite{Beck94}, who invoked
a small-scale dynamo at early times to kick-start the large-scale dynamo
at later times.

In galaxies, the turnover time on the integral scale can be as short
as $10^6$ or $10^7$ years, but with fluid Reynolds numbers well
above $10^{7}$, the relevant $e$-folding time of dynamo growth will
be shortened by a factor of $\Rey^{1/2}\approx3\times10^3$ or more.
Based on this argument, galactic dynamos may reach saturation first at 
the smallest eddy scales, on a
time scale as short as several hundred years.
A relevant limitation of reaching coherent large-scale fields comes
only from the late saturation phase when magnetic helicity fluxes are
expected to play an important role.

\section{Conclusions}
\label{Conclusions}

The results presented here support the
idea that large-scale fields can be efficiently generated 
even in the presence of strongly growing fluctuations driven by the
fluctuation dynamo. 
Clearly, the growth of the larger scale field is aided by the presence 
of helicity in the turbulence. But it is not as if there is an 
$\alpha^2$ large-scale dynamo independent of the small-scale dynamo;
as the growth rate of the $k=1$ mode does not seem to change
significantly right from the kinematic to the nonlinear stage. 
Rather it appears that there is one unified dynamo, with all
scales initially growing together at one rate, and then the largest scales
continuing to grow (aided by small-scale magnetic helicity loss) 
as the small-scale fields saturate. 

Several extensions of our model can be envisaged.
Our dynamo is a homogeneous one, making catastrophic
(resistive) quenching effects more pronounced.
It would therefore be useful to extend our studies to inhomogeneous
systems, for example when there is shear.
In that case, the magnetic energy density of the mean magnetic
field, $\meanBB^2$, is no longer constant in space, which leads to a
nonuniformity of the magnetic helicity flux divergence and can thereby
alleviate catastrophic (premature) quenching, as was shown by \cite{HB12}.
Such models should therefore be studied more thoroughly.
However, as astrophysical systems are all confined
in space with a corona and low-density material outside, it would be
useful to address such systems in some fashion.
It would be interesting to see whether this could remove the slow-down
of the growth caused by total magnetic helicity conservation
during the saturation phase.
This would be particularly important in view of understanding the observed
levels of coherent magnetic fields in young galaxies.

\section*{Acknowledgments}

We acknowledge the usage of the high performance computing facility
at IUCAA. PB acknowledges SRF support from CSIR, India
and currently from DOE, DE-FG02-12ER55142, at Princeton, USA.
This work was supported in part by
the Swedish Research Council grants No.\ 621-2011-5076 and 2012-5797,
as well as the Research Council of Norway under the FRINATEK grant 231444.

\bibliographystyle{mn2e}
\bibliography{ssdlsd}
\appendix

\label{lastpage}

\end{document}